\documentclass{nature}

\usepackage{graphicx}
\usepackage{multirow}
\graphicspath{ {images/} }
\DeclareGraphicsExtensions{.png,.pdf,.eps}

\usepackage{hhline}
\usepackage{epstopdf}
\usepackage{natbib}
\usepackage{amsmath,amssymb}

\usepackage[colorlinks,urlcolor=cyan,citecolor=blue,linkcolor=blue]{hyperref}

\usepackage{xcolor}

\newcommand{\notsotiny}{\fontsize{7.5}{7}\selectfont}

\title{Combined Exoplanet Mass and Atmospheric Characterization for Accelerated Exoplanetology}
\author{ 
Julien de Wit$^{1}$, 
Sara Seager$^{1,2,3}$, 
and Prajwal Niraula$^{1}$ 
} 

\begin{document}
\maketitle

\begin{affiliations}
\item Department of Earth, Atmospheric and Planetary Sciences, Massachusetts Institute of Technology, 77 Massachusetts Avenue, Cambridge, Massachusetts 02139, USA;
\item Department of Physics, Massachusetts Institute of Technology, 77 Massachusetts Avenue, Cambridge, MA 02139, USA;
\item Department of Aeronautics and Astronautics, Massachusetts Institute of Technology, 77 Massachusetts Avenue, Cambridge, MA 02139, USA
\end{affiliations}

\begin{abstract}

Today's most detailed characterization of exoplanet atmospheres is accessible via transit spectroscopy (TS). Detecting transiting exoplanets only yields their size, and it is thus standard to measure a planet's mass before moving towards their atmospheric characterization, or even the publication of their discovery. 
This framework, however, can act as a bottleneck for high-throughput exoplanetology. 
Here, we review existing applications of an alternative approach deriving exoplanet masses in small JWST atmospheric exploration programs and quantify the potential of its systematic application. 
We find that for $\sim$20\% of transiting exoplanets with existing mass constraints, a small JWST exploration program could yield the planetary mass with a similar--or better---precision. Such results suggest that proceeding directly with atmospheric exploration programs for favorable exoplanets (i.e., with a transmission spectroscopy metric, TSM, $\geq$100) could substantially reduce the time from detection to exoplanet atmospheric study and further support JWST's scientific output over its lifetime while saving up to 20\% of resources on radial-velocity (RV) facilities.
Furthermore, it can substantially increase the sample of characterized planets of three distinct subpopulations (Neptune-sized, young, and hot-star exoplanets), each providing specific insights into formation and evolution processes.
As the field of exoplanets increasingly turns to directly imaged planets, mastering the determination of planetary masses from atmospheric spectra will become essential.


\end{abstract}

\flushbottom
\maketitle


\newpage
\noindent \textbf{Towards an era of high-throughput, high-precision exoplanetology.}
\vspace{-7mm}

Two decades ago, only a handful of planetary candidates were discovered each year. The space-based missions Kepler \citep{Borucki2010} and TESS \citep{Ricker2015} increased that rate to hundreds annually, yielding thousands of transiting planets. While planet discovery from those missions has now plateaued, renewed high discovery rates are anticipated with the upcoming PLATO \citep{Rauer2014} and Roman \citep{Spergel2015} space missions---meaning the gap between candidate identification and confirmation will widen further.  With the new missions and with an ever-growing population of transiting exoplanets, the astronomical community will be poised with the prospects of in-depth ensemble studies of exoplanet atmospheres---primarily thanks to the James Webb Space Telescope \citep[JWST;][]{Gardner2006}. These two important changes in scale and scope translate into challenges and opportunities for the field. 

\noindent{\bf What's in a name?}

\vspace{-7mm}

One of the reasons we have seen such an increase in the number of known exoplanets is a cultural shift in what we call a planet \cite{etangs2022}. In the first decade or so of exoplanets being discovered by the transit method (i.e., pre-Kepler), researchers were adamant that an astronomical object {\it could only} be called a ``planet'' if verified by a mass measurement \cite{Sahu2006}. Brown dwarfs and binary star systems as well as ``noise'' (from stellar activity to instrument systematics)  all have ways of posing as planets,  and mass was thought to be the best way to rule out these false positives. Fast-forward to the era of Kepler, and now TESS, where our community happily accepts that a statistically ``validated'' planet may be deemed worthy of publication while lacking a mass measurement, thus illustrating a progressive shift in what we call a planet. The validation of Kepler's very first planet candidate (KOI-4.01, now Kepler-1658~b) required a whole decade and input from both radial-velocity and asteroseismology data  \cite{Chontos2019}, for example. While we still maintain a difference between ``confirmed'' (i.e., mass-measured) and ``validated'' (no mass but shown with high confidence to be of planetary nature), many studies and planet counts include validated planets.

The reason for the cultural shift away from requiring a confirmation is that if a planet is detectable by a transit survey, it is often easily detectable by many other observatories and can thus be ``validated''. 
However, it is not a given that a precise mass measurement (i.e., a ``confirmation'') will be easily within reach. While this is especially true for faint host stars or low-mass planets---because radial velocity requires dispersing starlight, standard (i.e., RV-based) mass measurements in general creates a bottleneck as RV instruments can only look at one star at a time whereas transit missions are monitoring hundreds of stars at once.

\vspace{3mm} \noindent{\bf Two textbook examples}
\vspace{-7mm}

The recent cases of the hot Jupiters WASP-193~b \citep{Barkaoui2024} and V1298\,Tau~b \citep{David2019} highlight two variants of the aforementioned bottleneck. 
Detected in 2015 amongst the WASP Survey data and validated shortly after with TRAPPIST-South, WASP-193~b required another 8 years before confirmation as a planet (and publication of its discovery), as constraints on its mass remained elusive for years. While the planet is a giant exoplanet ($R_p = 1.464 \pm 0.058~R_{\rm Jup}$), its unexpectedly-low density ($\sim$20$\times$ smaller than Jupiter's) delayed RV-based insights. Years after its detection, RV measurements finally yielded the first constraint on its mass ($M_p = 0.139 \pm 0.029 ~M_{\rm Jup}$, i.e., $\sim$20\% relative mass uncertainty---hereafter RMU). For V1298\,Tau~b, the issue in deriving a RV-based mass constraint was not a low planetary density, but rather high levels of stellar activity leading to stellar rotational modulation an order of magnitude above the expected planetary signal for this giant exoplanet. Recently, TS yielded V1298\,Tau~b's mass with a RMU below $10\%$ \citep{Blunt2023,Barat2024,Barat2025} and rejected at the $\sim$40$\sigma$
level of confidence the best fit
RV mass \cite{Suarez2022}. 
Other examples of TS-enabled mass measurements are presented in Refs.\cite{Alonso2021,Gandhi2023,Thao2024,Schmidt2025,Barat2025}, notably.


\vspace{3mm}\noindent {\bf A new synergy for mass and atmosphere measurements}
\vspace{-7mm}

\begin{figure}[!ht]
\hspace{-15mm}\includegraphics[width=1.2\textwidth]{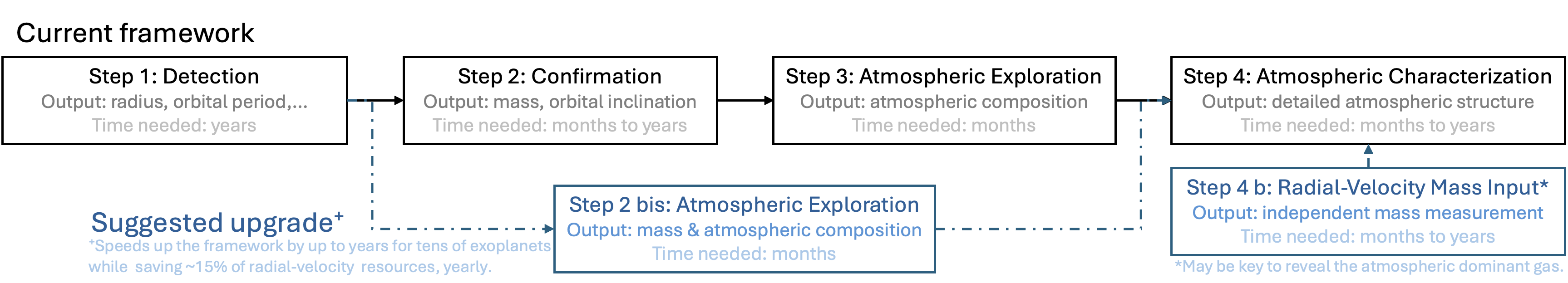}
\caption{\textbf{Flowchart of the exoplanet detection-to-characterization framework.} Current one-size-fits-all framework is shown in black, while the suggested upgrade is shown in blue. When applied to newly-detected exoplanets with TSM$\geq100$ (transmission-spectroscopy metric, Ref.~\cite{Kempton2018}), it speeds up the framework by months to years for tens of exoplanets while saving $\sim20\%$ of radial-velocity resources yearly (step 2 \textit{bis}). For a portion of these planets---primarily rocky worlds, a degeneracy between the planetary mass and the mean molecular weight ($\mu$) will require independent radial-velocity measurements to yield the mass and identify the primary atmospheric compound (step 4~b).}
\label{fig:figure1}
\end{figure}

We envision a new cultural shift for what is needed to consistently pass the bar from the discovery stage to the characterization stage (e.g., acquisition of JWST transmission spectra), similar to the one for our working definition of a planet. Such a cultural shift would save RV resources and speed up the atmospheric exploration of dozens of exoplanets. This cultural shift would translate into a revised detection-to-characterization framework to be applied to a subsample of new exoplanets (\autoref{fig:figure1}). This revised framework would benefit from a two-birds-one-stone approach, yielding preliminary atmospheric and mass constraints in one go, specifically from the few transit observations typical of JWST's atmospheric exploration programs.

\begin{figure}[!ht]
\hspace{-15mm}\includegraphics[width=1.15\textwidth]{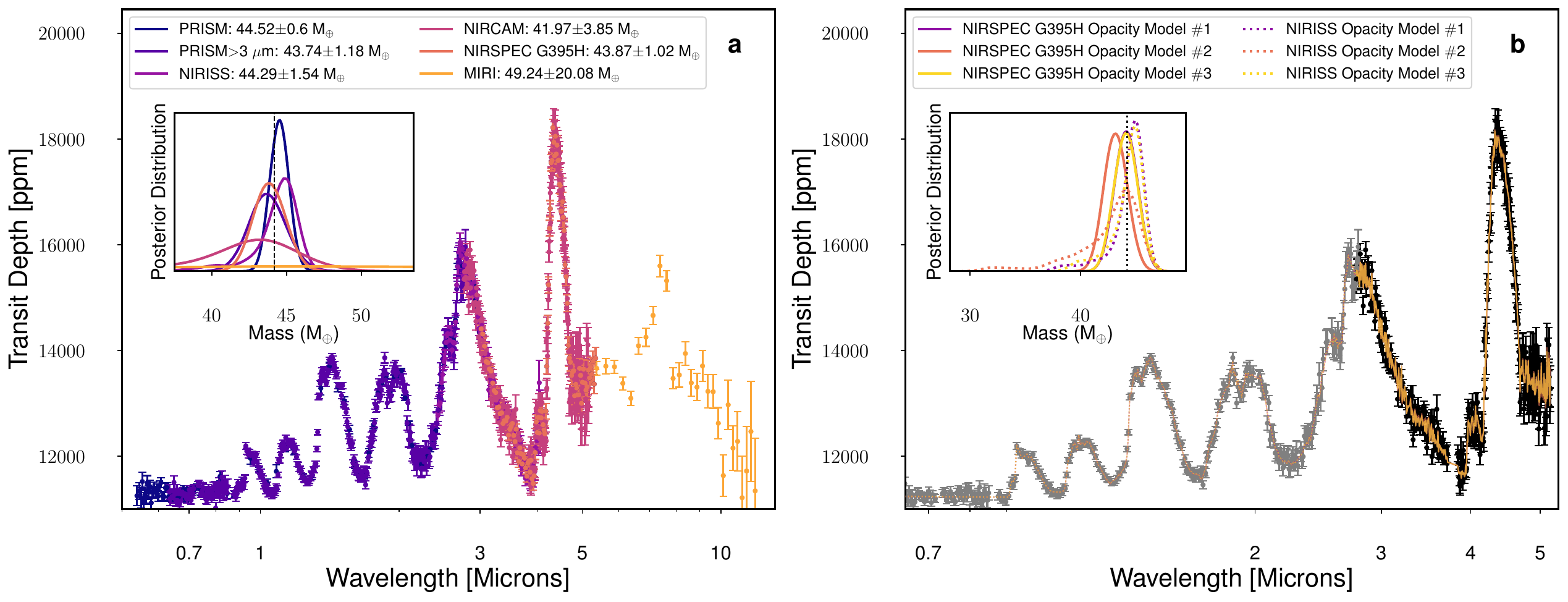}
\caption{\textbf{Exoplanet mass measurements via transmission spectroscopy.} \textbf{a.} Synthetic transmission spectrum of WASP-193~b-like planet when observed by various instruments aboard JWST. All instruments except MIRI allow constraining the mass with precision better than 10\%, with better than 1\% precision obtained with NIRSPEC/PRISM. The inset figure shows the posterior distribution of the mass obtained during the fit, where the black line shows the true mass used to create the synthetic spectrum. \textbf{b.} Same as left with a focus on testing the mass measurements sensitivity to imperfections of opacity models. Combined, these results show that mass measurements can be reliably performed despite stellar contamination, hazes, and/or imperfect opacity models for high-TSM atmospheres.}
\label{fig:figure2}
\end{figure}

Measuring exoplanet mass via transmission spectroscopy was proposed over a decade ago \cite{deWit2013}. Conceptually, information about the planetary mass is accessible through the atmospheric scale height ($H = k T/\mu g$---k is the Boltzmann constant, T is the temperature, $\mu$ is the molecular weight, and $g$ is the gravity), which is the primary driver for the amplitude of absorption features \cite{Miller2009}. Practically, the new-generation of observatories (esp. JWST) were needed to provide data with the resolution, coverage, and precision to yield independent constraints on its temperature, composition, and pressure profile (example presented in \autoref{fig:figure2}). 

\begin{figure}[!ht]
\centering\includegraphics[width=0.65\textwidth]{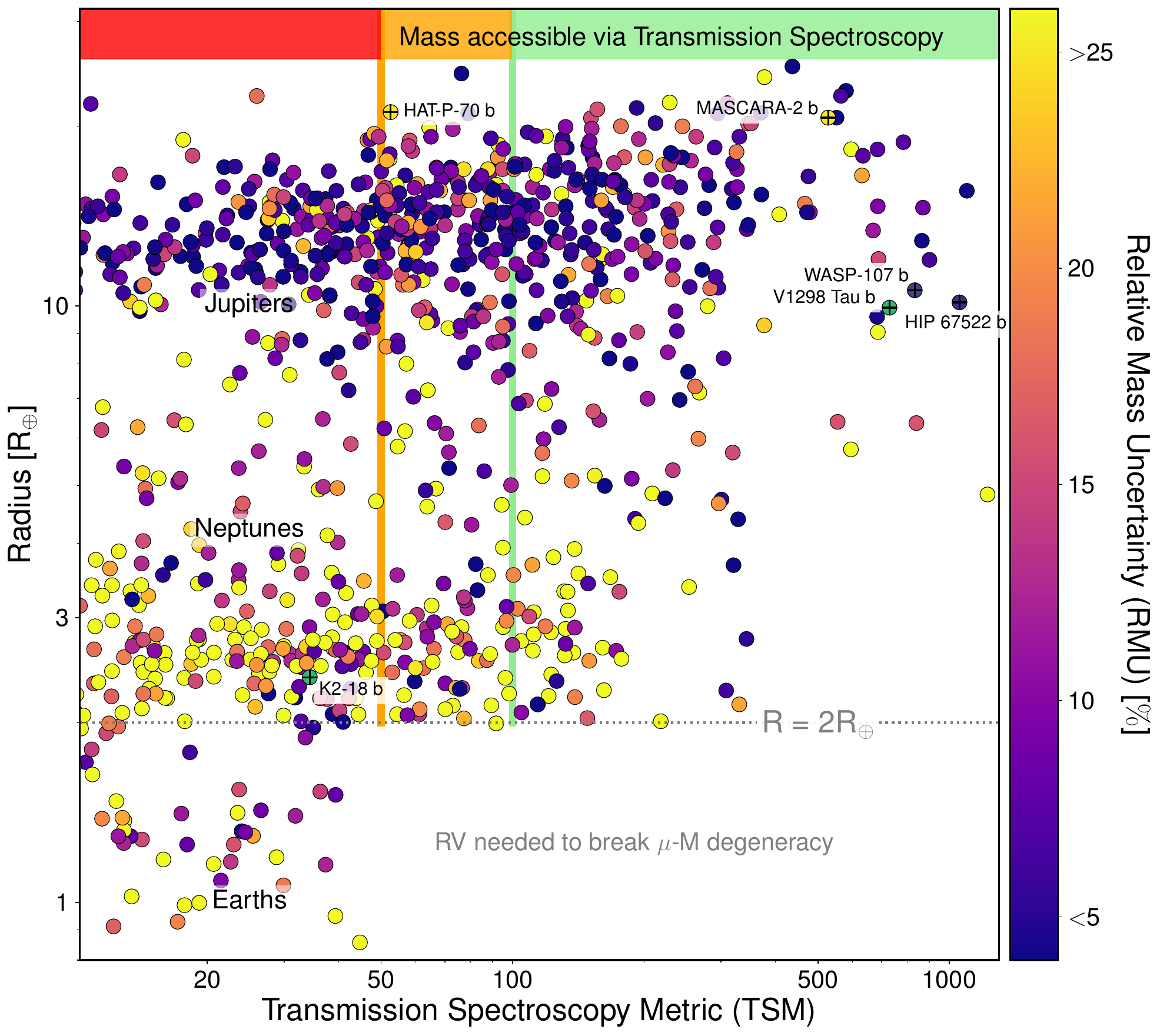}
\caption{\textbf{Application domain for mass measurements via small JWST atmospheric exploration programs.}  Size of known exoplanets with mass measurements (relative mass uncertainty, RMU, as color) versus their estimated TSM---data from Ref.\cite{exoplanet_eu}. Planets with mass informed by transmission spectroscopy are labeled (e.g., V1298~Tau\cite{Barat2024,Barat2025} and HIP~67522~b\cite{Thao2024}). For planets with TSM$\geq$100, direct mass measurements via atmospheric exploration is recommended---with a caveat for the rare terrestrial worlds with TSM$\geq$100 for which a degeneracy between the atmospheric mean molecular weight ($\mu$) and the planetary mass ($M$) may require independent mass insights from RV. 
While 734 exoplanets have 5$\sigma+$ mass constraints (i.e., RMU$<20\%$), 171 of them (23$\%$) reach a similar or better precision (RMU$\leq10\%$) with a small JWST exploration program ($\leq$2hrs of in-transit data). That number reaches 284 when accounting for all known exoplanets, which is equivalent to a third of the current mass-constrained population.   \label{fig:figure3}}
\end{figure}

The original study \cite{deWit2013} focused mostly on mass assessments for small planets, demonstrating notably the viability of such an alternative mass-measurement method for Earth-sized planets around M dwarfs within up to 40\,pc and super-Earths around K dwarfs (and later types) within up to 80\,pc with up to 200~hrs of JWST hours. Yet, several groups have recently shown that the technique is readily viable for some planets with much smaller JWST programs, with applications ranging from young hot worlds such as the Jupiter-size HIP~67522~b\cite{Thao2024} and Neptune-size V1298\,Tau~b \cite{Barat2024} to the temperate sub-Neptune K2-18~b \cite{Schmidt2025}\footnote{K2-18~b's mass was primarily constrained with radial-velocity data and recently refined with transmission-spectroscopy input from JWST.}---see \autoref{fig:figure3} for a revised application domain for small JWST exploration programs, with the planets listed above labeled. 


\noindent \textbf{A growing set of proof-of-concept applications}
\vspace{-7mm}

Though the technique was initially challenged \cite{Batalha2017}, the principal challenge to the adoption of TS-based mass measurements for exoplanets seems rooted in the lack of transmission spectra of sufficient quality, i.e., the delay of JWST's launch. Indeed, Ref.~\cite{Batalha2017}'s original findings were due to a focus on systems with low transmission spectroscopy metric \citep[TSM;][]{Kempton2018}---specifically super-Earths around Sun-like stars---and the same team later found that the mass of gas giants is within reach with single JWST visits to $\leq$10\% (Fig. 3; \cite{Batalha2019}). Subsequent studies in the 2020s have validated earlier findings on the viability of mass determination from transit transmission spectroscopy numerically \cite{Changeat2020, DiMaio2023} and via applied mathematical methods, such as transverse vector decomposition \citep{Matchev2022}. After a decade of numerical and theoretical validation, TS-based mass measurements are trickling in from independent teams for half a dozen of exoplanets \cite{Alonso2021,Gandhi2023,Thao2024,Barat2024,Schmidt2025,Barat2025}. 

\noindent \textbf{Updated performance tests}
\vspace{-7mm}

Over the last few years, the community has learned substantial amounts about the performances of new instruments (esp., JWST's) as well as new bottlenecks associated with TS. To properly make the case for TS-based mass measurements during small-scale atmospheric exploration programs, we performed performance tests accounting for our latest understanding of JWST's capabilities and TS-specific bottlenecks. 

We first used a synthetic transmission spectrum of WASP-193~b \cite{Barkaoui2024} to assess which JWST instrumental setup would yield the tightest mass constraints (see Methods). We find that NIRSPEC/PRISM, NIRSPEC/G395H, and NIRISS/SOSS perform best and can yield comparable performance (1-2\%---see \autoref{fig:figure2}, left panel) for such high-TSM ($\sim600$) planets. We also find that the information content gathered by NIRSPEC/PRISM beyond 3$\mu$m yields comparable constraints (at high TSM) to NIRISS and NIRSPEC/G395H, helping us address potential concerns associated with stellar contamination \citep{rackham2018, Rackham2023, Rackham2024} and hazes \citep{Fauchez2019,Komacek2020} as both potential bottlenecks are expected to have a marginal effect beyond 3$\mu$m \citep{Triaud2024}.

We then tested the impact of the opacity challenge \citep{Niraula2022} on mass measurements. To that end, we analyzed synthetic NIRISS and NIRSPEC/G395H spectra using 3 differently-perturbed cross-sections (see Methods). We find that the mass estimates are mostly insensitive to the simulated imperfections of opacity models. This is consistent with the findings of Ref.\cite{Niraula2023} that the opacity challenge does not currently affect the TS-based study of large, hot, and/or high-metallicity atmospheres (ie., most of the high-TSM atmospheres). 

Finally, we performed an ensemble of injection-retrieval analysis to understand the relationship between TSM and relative mass uncertainty (RMU hereafter) for a small exploratory JWST program (equivalent to 2 hours in transit, see Methods). We found that the RMU scales linearly with the TSM down to the TSM$\sim$75 regime; going from a RMU$\sim$1.5\% at TSM$\sim$600 to a RMU$\sim$4.5\% at TSM$\sim$200 and $\sim$9\% at TSM$\sim$100, to finally diverge to $\sim$25\% between TSM$\sim$75 and TSM$\sim$50 depending on the planetary properties.

\noindent \textbf{Expectations vs reality: existing applications plagued by flares, partial data, or high clouds}
\vspace{-7mm}

Owing to their high TSMs, we expect RMUs below 2\% from the small JWST programs behind the mass measurements of V1298~Tau~b (1 NIRSPEC/G395H transit via GO-2149, P.I. Désert), HIP~67522~b (1 NIRSPEC/G395H transit via GO-2498, P.I. Mann), and WASP-107~b (1 NIRSPEC/G395H transit via GTO-1224, P.I. Birkmann). Instead, applications found in the literature report RMUs between 7 and 10\% \cite{Thao2024,Barat2025}. We explain below this tension between expectations and reality by a selection bias that artificially diminishes the application domain of TS-based mass measurements, if taken at face value. 

The transmission spectroscopy metric is a theoretical construct that does not account for practical realities such as the effect of stellar activity, suboptimal observations, or high-altitude clouds. Such extra noise/dilution sources can cause tension between expectations and reality. To quantify that aspect, we note that a TSM$\sim$600 planet should yield an amplitude-to-errorbar ratio of $\sim$30 for CO$_2$'s 4.3-$\mu$m feature for an exploratory program with NIRSPEC/G395H (see \autoref{fig:figure2}), yet the observed ratio is $\sim$10 for V1298~Tau~b \cite{Barat2025}. The 3$\times$ drop in signal-to-noise ratio (SNR) is driven by an intense flare (amplitude equivalent to $\sim$80$\%$ the transit depth) covering two-thirds of the transit duration. For HIP~67522~b, a similar 3$\times$ SNR drop is observed, this time due to parts of the standard transit data missing (incl., the entire egress and post-transit baseline) \cite{Thao2024}. Finally, for WASP-107~b, clouds/hazes appear to be limiting the range of pressure TS-accessible to 0.01-1mbar \cite{Welbanks2024}. In comparison, the first comparative study of hot Jupiters reported pressure levels down to $\sim$100mbar probed consistently in transmission \cite{Sing2016}. Reducing the maximum pressure level probed from $\sim$100mbar to $\sim$1mbar corresponds to a reduction in the extent of the atmospheric annulus probed by $\sim4.5H$, to compare to a typical annulus extent of $\sim7H$ \cite{Miller2009,deWit2013} (i.e., a SNR drop of 3$\times$).

In light of this context, we expect TS-based measurements to perform better than currently showcased in the aforementioned, promising applications---i.e., TS-based RMUs closer to our predictions than to those associated with young planets around stars so active that their masses are beyond RV's reach, or of a planet with a remarkably-high cloud deck. Moving forward, we therefore use the TSM cutoff presented in the previous section; RMU$\sim$10\% for TSM$\sim$100.

\noindent \textbf{The exoplanet population ripe for cheap TS-based mass measurements}
\vspace{-7mm}

\begin{figure}[!ht]
\vspace{-30mm}
\includegraphics[width=1.\textwidth]{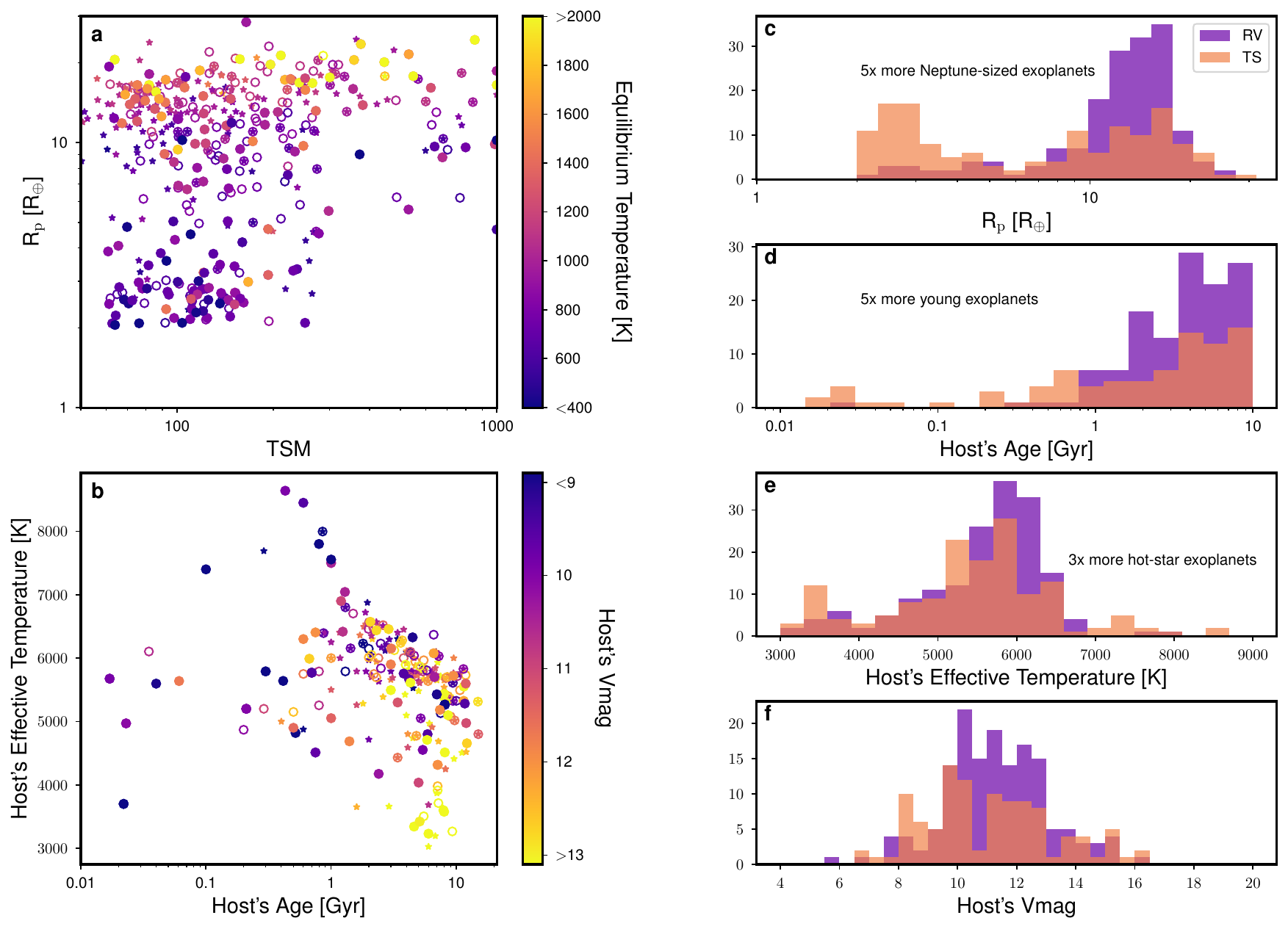}
\caption{\textbf{The exoplanet population ripe for mass measurements via transmission spectroscopy.}  
\textbf{a.} Size of known exoplanets with RMU $\leq10\%$ (or expected to be such with TS) versus their estimated TSM with equilibrium temperature as color---data from Ref.\cite{exoplanet_eu}. Planets with RV-based mass-measurements shown as stars, for which TS is expected to perform better than RV as empty circles, for which TS-based mass is within reach while no RV-based mass exist as filled circles---see \autoref{tab:Table1}. \textbf{b.} Same sample as \textbf{a.} but showing properties of the exoplanets' host stars. \textbf{c.} Histogram of exoplanet size for the sample with existing RV-based masses and the sample without mass but within reach of TS. \textbf{d.} Same as \textbf{c.} but for the exoplanets' host-star age. \textbf{e.} Same as \textbf{c.} but for the exoplanets' host-star effective temperature. \textbf{f.} Same as \textbf{c.} but for the exoplanets' host-star V magnitude. It shows that three distinct subpopulations may particularly benefit from this new framework, thereby consolidating the insights to be gained via their study into formation and evolution processes.
\label{fig:figure4}}
\end{figure}

Not all exoplanets are equal, nor are the facilities accessible to the community. We leveraged our findings above to identify the population of exoplanets detected readily amenable for TS-based mass measurements (\autoref{fig:figure3}). We found that while 734 exoplanets have significant RV-based mass constraints (5$\sigma+$ measurement, i.e., RMU$<20\%$ with data gathered from Ref.\cite{exoplanet_eu}), 171 of them (23$\%$) would reach a similar or better precision (RMU$\lesssim10\%$) with a small JWST exploration program ($\leq$2hrs of in-transit data). That number reaches 284 when accounting for all known exoplanets, which is equivalent to a third of the current mass-constrained population. Out of these 284 exoplanets, 150 do not have significant RV-based mass measurements. This means that a TS initiative could increase the sample of exoplanets with informative mass constraints by up to $20\%$. We provide the list of these 150 exoplanets favored for such an initiative in \autoref{tab:Table1} and showcase them in \autoref{fig:figure4}.

The implementation of this new framework can substantially increase the sample of characterized planets of three distinct subpopulations (see \autoref{fig:figure4}). First, Neptune-sized exoplanets suitable for atmospheric studies; while 13 of such exoplanets are currently with RMUs$\leq10\%$, TS could add 52 new ones---i.e., a 5$\times$ sample increase. Second, planets around young stars with significant stellar variability as captured by the V1298~Tau~b example provided early on; 4 with RMUs$\leq10\%$, 21 within TS reach---i.e., a 5$\times$ sample increase. Third, planets around hot and/or rapidly-rotating stars for which fewer absorption lines exist and those that do exist are rotationally broadened. This notably affects surveys such as the Earth Twin Survey \cite{Gupta2021} for which a large fraction of stars in the nominal sample are above the Kraft break and alternative mass measurements are needed and considered, such as astrometry \cite{Giovinazzi2025}. For this subpopulation of hot-star planets, we anticipate a 3$\times$ sample increase; 4 with RMUs$\leq10\%$, 11 within TS reach.  

\noindent \textbf{RV-enhanced atmospheric characterization}
\vspace{-7mm}

On the other end of the application domain, we identified two groups of exoplanets that will reliably benefit from RV-based mass measurements: (1) those for which an informative constraints on the planetary mass with a few JWST transits is not viable (TSM$<<$50) and (2) those for which the dominant atmospheric constituent cannot be identified from existing transmission measurements (typically sub-Neptunes, see \autoref{fig:figure3}, left panel). In the first case, we recommend performing the RV-based mass assessment ahead of the atmospheric exploration to refine the planet's TSM and assess if it is worth pursuing as a target for JWST investigations. In some instances (see previous section and Ref.\cite{TJCI2024}), the planetary mass may remain beyond reach for RV and the target may be deemed of high-enough interest to warrant proceeding with $\mathcal{O}$(100) JWST hrs to proceed with a detailed atmospheric characterization while attempting to reach a mass constraint (e.g., for a terrestrial temperate world around a late M dwarf \cite{deWit2013}). Planets with TSMs down to $\mathcal{O}$(10) may then be accessible to TS-based mass measurements.

In the second case, atmospheric characterization is already underway, and a strong degeneracy exists between the atmospheric mean molecular weight ($\mu$) and the planet's mass. Specifically, the same observed scale height can be explained by either a low-mass planet with a high-$\mu$ atmosphere or a high-mass planet with a low-$\mu$ atmosphere, especially when the dominant background gas lacks detectable spectral features \cite{Niraula2025, deWit2025}. For example, as metallicity increases, it becomes more challenging to identify the signature of H$_2$-H$_2$ collision-induced absorption \cite{deWit2025}, and thus a degeneracy may arise as follows: the background gas is not a strong absorber, is it H$_2$, N$_2$, or something else? It results in a bimodal distribution on $\mu$ peaking around 2.3 and 28 results in a bimodal distribution on the planetary mass with values similarly spread apart by an order of magnitude---the same goes for disentangling between, e.g., H$_2$O- ($\mu\sim18$) and CO$_2$-rich ($\mu\sim40$) atmospheres. Such a degeneracy that can be broken by a loose (i.e., a factor of a few) constraint on the planetary mass from RV measurements or, possibly, from interior models \cite{deWit2025}.

\noindent \textbf{Immediate impacts on scientific output}
\vspace{-7mm}

Each year that passes corresponds to opportunity losses for our facilities in two ways. The first is mentioned above for RV facilities: for $\sim$20\% of the exoplanets monitored, RV facilities yield mass constraints that could eventually be provided by JWST observations if/when their atmospheric exploration starts. The second relates to JWST and its finite lifespan: high-value exoplanets like WASP-193~b may await mass measurements for years and barely make it to JWST's target lists on time to support their in-depth characterization---if anything exotic is revealed in their atmospheres warranting additional observations after the exploration stage. Quantitatively speaking of our first textbook example: while RV measurements only constrained WASP-193~b's mass to within $\sim$20\% so far, JWST's Cycle 4 transit observations (GO~9101, PI: Radica) could yield its mass to within $\sim$1\%. Other top targets are listed in \autoref{tab:Table1} ahead of the deadline for JWST's Cycle 5 Call for Proposals.

\noindent \textbf{Paving the way to the era of directly-imaged exoplanets}
\vspace{-7mm}

\begin{figure*}[!ht]
\centering
\vspace{-3mm}\includegraphics[width=0.65\textwidth]{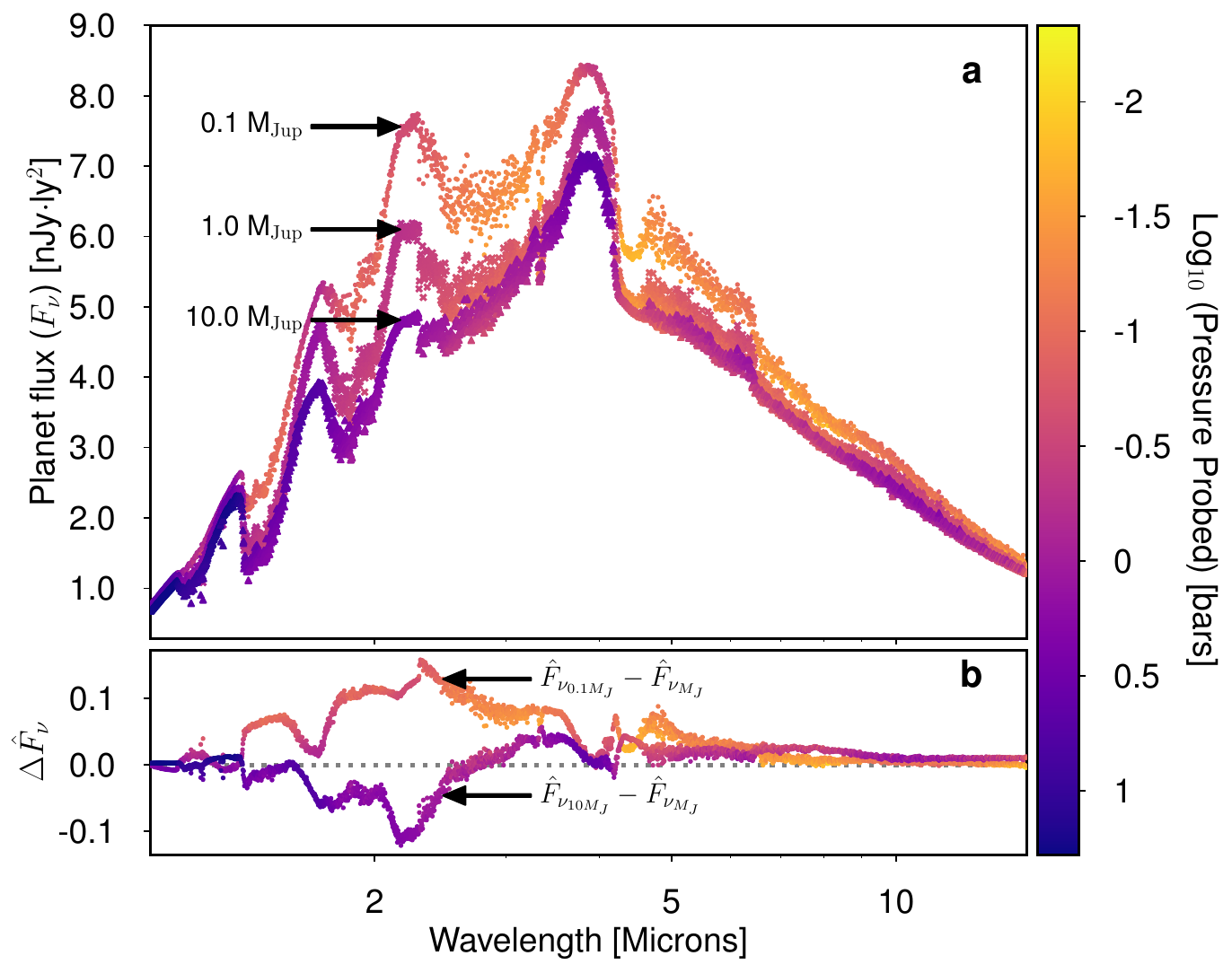}
\caption{\textbf{Towards mass measurements from emission spectroscopy.} Synthetic emission spectra of a Jupiter-sized planet for 3 different masses (0.1, 1, and 10 M$_{\rm Jup}$), all other properties unchanged. It highlights how the mass of a planet shapes its emission spectrum---esp. the amplitude of absorption features---via the density gradient along the light path; the lower the planetary mass, the larger atmospheric scale height, the larger absorption features, and the shallower the pressure levels probed.
As size-brightness degeneracies will exist, the bottom panel presents the difference between the normalized flux ($\Delta\hat{F_\nu}$). 
Information about a planet's mass is thus recorded in its emission spectrum, which will be of relevance for future directly-imaged planets---most of which will be inaccessible with RV facilities.
\label{fig:figure5}}
\end{figure*}

We show in \autoref{fig:figure5} how the basics introduced in Ref.\,\cite{deWit2013} for transmission spectroscopy are likely to hold for emission spectroscopy---just as they do for stars and brown dwarfs for which mass (or gravity) measurements have been performed for decades using emission spectra \cite{Kleinmann1986,Greene1995,McGovern2004}. \autoref{fig:figure5} highlights how the mass of a planet shapes its emission spectrum---esp. the amplitude of absorption features (bottom panel)---via the density gradient along the light path. As for TS, a lower planetary mass yields a larger atmospheric scale height and thus larger absorption features, while preventing deep pressure levels to be probed. While a detailed exploration of this technique's application domain is beyond this Perspective, we can state that information about a planet's mass is recorded in its emission spectrum. This will be of particular relevance for future directly-imaged planets---many of which will be inaccessible with RV facilities due to orbits with high inclination and/or period.

We thus argue to conclude this Perspective that diligently working towards practicing exoplanet mass measurements from their atmospheric spectra now will not only boost our immediate scientific output but also support the next generation of exoplaneteers and observatories to come such as the Habitable World Observatory \cite{Gaudi2020} and the Nautilus Space Observatory \cite{Apai2022}.



\vspace{0.5cm}

\newpage
\pagebreak
\clearpage

\section{Figures}

\newpage
\pagebreak
\clearpage

\begin{table}
    
    \vspace{-25mm}\hspace{-30mm}
    \notsotiny
    \begin{tabular}{r|l|r|l|r|l|r|l|r|l}
     $^{+}$AU Mic b$^{NY}$ (\underline{1000}) & 1;30 & HD 93963 Ac$^{N}$ (193) & 5;N/A & TIC 290048573 b$^{N}$ (120) & 8;N/A & TOI-1803 c$^{N}$ (160) & 6;47 & TOI-4438 b$^{N}$ (136) & 7;21 \\ 
AU Mic c$^{NY}$ (71) & 13;23 & HIP 41378 e$^{N}$ (70) & 13;42 & TIC 303682623 b$^{}$ (234) & 4;N/A & TOI-2018 b$^{N}$ (64) & 14;23 & TOI-4468 b$^{}$ (122) & 7;N/A \\ 
DS Tuc Ab$^{Y}$ (530) & 2;\underline{50} & HIP 41378 f$^{}$ (372) & 2;\underline{24} & TIC 434398831 c$^{Y}$ (298) & 3;\underline{15} & $^{*}$TOI-2031 Ab$^{}$ (71) & 13;29 & TOI-469 b$^{N}$ (151) & 6;32 \\ 
$^{+}$GJ 3090 b$^{N}$ (251) & 4;22 & $^{+}$HIP 67522 b$^{Y}$ (987) & 1;\underline{7} & TOI-1064 c$^{N}$ (141) & 6;74 & $^{+}$TOI-2076 b$^{NY}$ (162) & 6;\underline{53} & TOI-5082 b$^{N}$ (151) & 6;N/A \\ 
HAT-P-57 Ab$^{H}$ (86) & 10;50 & HIP 67522 c$^{Y}$ (116) & 8;N/A & TOI-1097 b$^{NY}$ (126) & 7;N/A & $^{+}$TOI-2076 c$^{NY}$ (231) & 4;\underline{50} & TOI-5143 Ac$^{}$ (356) & 3;N/A \\ 
HAT-P-64 b$^{}$ (118) & 8;27 & K2-138 f$^{N}$ (113) & 8;\underline{104} & TOI-1135 b$^{}$ (676) & 1;48 & $^{+}$TOI-2076 d$^{NY}$ (122) & 7;\underline{38} & TOI-554 b$^{N}$ (111) & 8;31 \\ 
HAT-P-67 Ab$^{}$ (376) & 2;33 & K2-266 Ab$^{N}$ (121) & 7;\underline{80} & TOI-1136 d$^{NY}$ (278) & 3;20 & TOI-2120 b$^{N}$ (76) & 12;50 & TOI-622 b$^{}$ (93) & 10;24 \\ 
HAT-P-70 b$^{YH}$ (64) & 14;50 & K2-308 b$^{}$ (249) & 4;50 & TOI-1136 f$^{NY}$ (130) & 7;32 & TOI-2128 Ab$^{N}$ (157) & 6;42 & TOI-6478 b$^{N}$ (110) & 8;N/A \\ 
HATS-27 b$^{}$ (75) & 12;25 & K2-309 Ab$^{Y}$ (639) & 1;\underline{19} & TOI-1184 b$^{N}$ (111) & 8;34 & TOI-2136 b$^{N}$ (64) & 14;61 & TOI-6883 Ac$^{}$ (140) & 6;\underline{37} \\ 
HATS-37 Ab$^{}$ (108) & 8;42 & K2-384 f$^{N}$ (124) & 7;N/A & $^{+}$TOI-1231 b$^{N}$ (93) & 10;21 & TOI-2154 b$^{}$ (81) & 11;21 & TOI-7018 b$^{}$ (106) & 8;N/A \\ 
HATS-39 b$^{}$ (92) & 10;21 & KELT-15 Ab$^{}$ (106) & 9;32 & TOI-1246 Ad$^{N}$ (82) & 11;32 & TOI-216 b$^{}$ (222) & 4;\underline{4} & $^{+}$TOI-836 c$^{N}$ (72) & 12;28 \\ 
HATS-43 b$^{}$ (188) & 5;21 & KELT-17 b$^{}$ (265) & 3;22 & TOI-1247 b$^{N}$ (61) & 15;37 & TOI-2364 b$^{}$ (98) & 9;22 & TOI-905 Ab$^{}$ (121) & 7;\underline{6} \\ 
HATS-46 b$^{}$ (123) & 7;36 & KELT-20 b$^{Y}$ (445) & 2;50 & $^{+}$TOI-125 b$^{N}$ (116) & 8;\underline{9} & TOI-2443 b$^{N}$ (121) & 7;27 & TOI-907 Ab$^{}$ (101) & 9;N/A \\ 
HATS-47 b$^{}$ (100) & 9;\underline{7} & Kepler-105 b$^{N}$ (193) & 5;\underline{102} & $^{+}$TOI-125 c$^{N}$ (137) & 7;\underline{15} & TOI-257 b$^{}$ (101) & 9;\underline{17} & WASP-102 b$^{Y}$ (88) & 10;21 \\ 
HATS-62 b$^{}$ (143) & 6;50 & Kepler-139 d$^{}$ (\underline{1000}) & 1;41 & TOI-1422 b$^{N}$ (61) & 15;21 & TOI-2583 Ab$^{}$ (92) & 10;23 & WASP-121 b$^{}$ (360) & 2;\underline{5} \\ 
HATS-71 b$^{}$ (148) & 6;53 & Kepler-447 b$^{}$ (82) & 11;34 & TOI-1453 Ac$^{N}$ (98) & 9;27 & TOI-2587 Ab$^{}$ (67) & 13;23 & WASP-122 Ab$^{}$ (196) & 5;\underline{5} \\ 
$^{+}$HATS-72 b$^{}$ (143) & 6;\underline{3} & Kepler-450 Ac$^{}$ (165) & 5;N/A & TOI-1471 b$^{N}$ (89) & 10;32 & TOI-261 b$^{N}$ (171) & 5;N/A & WASP-145 Ab$^{Y}$ (114) & 8;\underline{4} \\ 
$^{+}$HD 106315 c$^{N}$ (98) & 9;32 & LHS 1678 b$^{N}$ (145) & 6;129 & TOI-1472 b$^{N}$ (66) & 14;31 & TOI-2669 b$^{}$ (61) & 15;31 & WASP-167 b$^{H}$ (548) & 2;\underline{81} \\ 
HD 110067 b$^{N}$ (110) & 8;32 & MASCARA-1 b$^{H}$ (82) & 11;24 & TOI-1518 b$^{H}$ (157) & 6;26 & TOI-270 c$^{N}$ (115) & 8;\underline{5} & WASP-172 b$^{H}$ (217) & 4;21 \\ 
HD 110067 c$^{N}$ (111) & 8;50 & MASCARA-4 Ab$^{H}$ (233) & 4;\underline{14} & TOI-1669 b$^{N}$ (93) & 10;64 & $^{+}$TOI-270 d$^{N}$ (104) & 9;\underline{4} & WASP-174 b$^{}$ (257) & 4;28 \\ 
HD 110067 d$^{N}$ (120) & 7;39 & NGTS-26 b$^{}$ (74) & 12;22 & TOI-1683 b$^{N}$ (142) & 6;\underline{24} & TOI-2796 b$^{}$ (251) & 4;23 & $^{+}$WASP-178 b$^{YH}$ (211) & 4;33 \\ 
HD 110067 f$^{N}$ (125) & 7;38 & Qatar-6 Ab$^{}$ (188) & 5;\underline{10} & TOI-1710 Ab$^{N}$ (144) & 6;26 & TOI-2818 b$^{}$ (121) & 7;37 & $^{*}$WASP-193 Ab$^{}$ (622) & 1;22 \\ 
HD 110067 g$^{N}$ (68) & 13;50 & SOI-1 b$^{H}$ (\underline{1000}) & 1;N/A & TOI-1716 b$^{N}$ (155) & 6;44 & TOI-2876 b$^{}$ (157) & 6;31 & WASP-194 Ab$^{Y}$ (69) & 13;23 \\ 
HD 183579 b$^{N}$ (136) & 7;242 & SOI-2 b$^{H}$ (854) & 1;N/A & TOI-1723 b$^{N}$ (76) & 12;48 & TOI-3135 b$^{}$ (85) & 11;28 & WASP-195 b$^{Y}$ (172) & 5;29 \\ 
HD 191939 d$^{N}$ (101) & 9;50 & SOI-7 b$^{H}$ (531) & 2;N/A & TOI-1749 Ac$^{N}$ (63) & 14;173 & TOI-3474 b$^{}$ (79) & 11;24 & WASP-33 Ab$^{YH}$ (300) & 3;\underline{19} \\ 
$^{+}$HD 207496 Ab$^{NY}$ (147) & 6;26 & TIC 12999193 b$^{}$ (793) & 1;N/A & TOI-1758 b$^{N}$ (237) & 4;91 & TOI-3682 Ab$^{}$ (89) & 10;36 & WASP-34 b$^{}$ (381) & 2;155 \\ 
HD 63433 c$^{NY}$ (94) & 10;27 & TIC 165227846 b$^{}$ (880) & 1;N/A & TOI-1759 Ab$^{N}$ (117) & 8;28 & TOI-3693 b$^{}$ (64) & 14;24 & WASP-49 Ab$^{}$ (272) & 3;\underline{8} \\ 
HD 63935 c$^{N}$ (66) & 14;22 & TIC 251090642 b$^{}$ (104) & 9;N/A & $^{+}$TOI-178 d$^{N}$ (111) & 8;27 & TOI-3785 b$^{N}$ (97) & 9;27 & WASP-55 Ab$^{}$ (147) & 6;21 \\ 
HD 73344 b$^{N}$ (78) & 12;50 & TIC 262605715 b$^{}$ (145) & 6;N/A & $^{+}$TOI-178 g$^{N}$ (80) & 11;37 & TOI-3976 b$^{}$ (75) & 12;21 & WASP-79 b$^{}$ (226) & 4;\underline{10} \\ 
HD 73583 Ab$^{NY}$ (93) & 10;32 & TIC 268727719 b$^{}$ (121) & 7;N/A & TOI-1798 b$^{N}$ (151) & 6;29 & TOI-4145 Ab$^{}$ (207) & 4;30 & $^{+}$WD 1856+534 Ab$^{}$ (\underline{1000}) & 1;N/A \\  
    \end{tabular}
    \caption{{\bf Top 150 exoplanets for TS-based mass measurements.} TSM in parenthesis next to name, together with  the TS- and current RMU in percent in column next to each name, separated by semicolon. Underlined current RMUs point to masses measured with a different technique than RV. Exoplanets with a $^{+}$ or $^{*}$ already have JWST observations executed or pending \cite{Nikolov2022}, respectively. Exoplanets with a $^{N}$ or $^{Y}$ or $^{H}$ belong to the TS-enhanced subpopulations of Neptune-sized exoplanets, young exoplanets, or hot-star exoplanets, respectively.}
    \label{tab:Table1}
\end{table}

\newpage
\pagebreak
\clearpage

\begin{addendum}
 
\item[Author Contributions] 

J.d.W. designed and led the study with support from S.S. and P.N. P.N. performed the simulations behind Figs. 2 to 4 with support of J.d.W.

\item[Acknowledgments]
The authors thank Daniel Apai, Saugata Barat, Jennifer Burt, René Doyon,  Benjamin Rackham,  and Jeff Valenti for their valuable inputs. The authors acknowledge support from the European Research Council (ERC) Synergy Grant under the European Union’s Horizon 2020 research and innovation program (grant No. 101118581). The contributions of J.d.W. and S.S. were funded, in part, by the Heising-Simons Foundation through grant No. 2024-5688. This research has made use of data obtained from or tools provided by the portal exoplanet.eu of The Extrasolar
Planets Encyclopaedia \cite{exoplanet_eu}.

\item[Competing Interests] The authors declare that they have no competing financial interests.

\item[Data Availability]

n/a

\item[Code Availability]

 \item[Correspondence] Correspondence and requests for materials
should be addressed to J.d.W~(email: jdewit@mit.edu). 

\end{addendum}

\newpage
\pagebreak
\clearpage


\section{Methods}

\subsection{Performance Tests}

To revisit the expected performance of mass retrieval using transmission spectroscopy in a context of improved constraints on instrument performances (and new bottlenecks such the stellar-contamination \cite{rackham2018} and opacity challenges \cite{Niraula2022}), we generated an ensemble of synthetic transmission spectra and perform sets of retrievals following the procedures used in Refs.\cite{Niraula2022,Niraula2025,Niraula2025} with \texttt{tierra}\footnote{\texttt{tierra} is publicly available at \href{https://github.com/disruptiveplanets/tierra}{https://github.com/disruptiveplanets/tierra}}. We first generated synthetic transmission spectra of WASP-193~b like planet testing the performance of various JWST instruments, and identified NIRISS and NIRSPEC/PRISM as the best instruments for constraining the mass. To that end, we model the synthetic planet using physical properties of WASP-193 b \citep{Barkaoui2024} with an isothermal temperature of 1252 K and four absorbers: water, carbon dioxide, carbon monoxide, and sulfur dioxide with abundances of 1000 ppm, 100 ppm, 10 ppm, and 100 ppm respectively and the rest consisting of hydrogen and helium at a ratio of 85/15. As for the noise, we assumed the same noise levels as were observed for WASP-39~b's across its ERS observations (GO 1366; PI: Natalie Batalha), typically in the range of 100-200 ppm. (Note that although the TSM metric uses planet mass, it can be estimated from the radius alone \citep{Chen2017,Kempton2018}.)

To test the effect of imperfection in opacity models on mass measurements, we focused on two instruments (NIRISS and NIRSPEC/PRISM) and performed cross-retrievals on the aforementioned synthetic spectrum with three versions of the perturbed cross-sections, representing perturbations on pressure broadening parameters (using air broadening \texttt{CS-N2-25}, and self broadening \texttt{CS-SELF}) as well as line wings (extended to 500 Half width Half Maximum - \texttt{CS-N2-500}) as in \cite{Niraula2023}.

\subsection{Emission Spectra}
To assess the sensitivity of exoplanets' emission spectra to their masses, we generated self-consistent p-T profile using Guillot Profile \citep{guillot2010} assuming internal temperature of 100 K, equilibrium temperature of 1500 K,  mean infrared opacity ($\kappa$) of 0.1 and the ratio of visual to infrared opacity ($\gamma$) set at 0.4; only varying the mass from 0.1 to 10 M$_{\rm Jup}$. The p-T profiles are then fed into \textit{petitRadtrans} \citep{molliere2019} to generate the emission spectra presented in \autoref{fig:figure5}, assuming the same atmospheric composition as used for transmission spectra. The clear differences between spectra for planets varying only in their masses show that emission spectra (just like transmission spectra) record the planetary-mass information through the atmospheric scale height. A lower planetary mass yields a larger atmospheric scale height and thus lower pressure levels probed, for example. \\

\section*{References}
\bibliography{Refs_All.bib}
\bibliographystyle{naturemag}




\end{document}